\documentclass[showpacs,aps,prd,superscriptaddress,nofootinbib,floatfix,amsmath,amssymb]{revtex4}
\usepackage{graphicx}
\usepackage{amssymb}

\begin{document}


\title{Static quantities of a neutral bilepton in the 331 model with right-handed neutrinos}
\author{G. Tavares-Velasco}
\email[E-mail:]{gtv@fcfm.buap.mx}
\author{J. J. Toscano}
\email[E-mail:]{jtoscano@fcfm.buap.mx} \affiliation{Facultad de
Ciencias F\'\i sico Matem\'aticas, Benem\'erita Universidad
Aut\'onoma de Puebla, Apartado Postal 1152, Puebla, Pue.,
M\'exico}

\date{\today}

\begin{abstract}
A neutral vector boson can possess static electromagnetic
properties provided that the associated field is no
self-conjugate. This possibility is explored in the $SU_C(3)
\times SU_L(3)\times U_N(1)$ model with right-handed neutrinos,
which predicts a complex neutral gauge boson $Y^0$ in a nontrivial
representation of the electroweak group. In this model the only
nonvanishing form factors are the CP-even ones, which arise from
both the quark and gauge sectors, and contribute to the magnetic
dipole and the electric quadrupole moments of this neutral
particle.
\end{abstract}


\maketitle

\section{Introduction}
\label{int} The electromagnetic properties of neutral particles
have been the source of great interest since they are generated at
the loop level, thereby opening up the possibility for the
detection of new physics effects. Considerable attention has been
paid to the electromagnetic properties of neutrinos and the
neutral $Z$ boson of the standard model (SM). In particular, the
impact of new physics effects on the trilinear couplings of the
$Z$ boson has been studied in a model-independent manner using the
effective Lagrangian technique \cite{Belanger}. As far as neutral
fermions are concerned, it was long realized that the off-shell
electromagnetic vertex of a massless Dirac neutrino is a
gauge-dependent quantity \cite{Marciano}. On the other hand, a
massive Dirac neutrino does have static electromagnetic properties
which characterize its magnetic and electric dipole moments. This
is to be contrasted with the case of a Majorana neutrino, which
only has off-shell electromagnetic properties \cite{Majorana},
which in turn is a consequence of the fact that a Majorana
neutrino is identical to its antiparticle. A more recent
model-independent study of the electromagnetic form factors of
Majorana particles with higher spin was presented in Ref.
\cite{Boudjema}. The situation for neutral spin-1 particles is
similar as for neutrinos: a neutral vector boson characterized by
a self-conjugate field, for which the particle is identical to its
antiparticle, cannot have static electromagnetic properties. This
fact has been already discussed in the case of the neutral $Z$
boson \cite{Barroso}. On the contrary, a no self-conjugate field
do can have static electromagnetic properties.

The possibility that neutral particles have nonzero static
electromagnetic properties was explored in a general context using
arguments of gauge invariance and transformation under the
discrete symmetries $C$, $P$ and $T$ \cite{nieves}. Several
extensions of the SM, such as grand unified theories (GUTs),
predict the existence of at least one new complex neutral gauge
boson with nonzero content of quantum numbers from the global or
local symmetries of the theory. The purpose of this work is to
present a calculation in a specific version of the $331$ model
\cite{Fra-Ple} which predicts the existence of a no self-conjugate
neutral gauge boson in a nontrivial representation of the
electroweak group.

The $331$ model is based on the simplest non-Abelian extension of
the SM group, namely, $SU_c(3)\times SU_L(3)\times U_N(1)$
\cite{Fra-Ple}. This model is appealing and has been the source of
interest recently \cite{Tavares-331} because it requires that the
number of fermion families be a multiple of the quark color number
in order to cancel anomalies, which suggest a path to the solution
of the flavor problem. Another important feature of this model is
that the $SU_L(2)$ group is totally embedded in $SU_L(3)$. As a
consequence, after the first stage of spontaneous symmetry
breaking (SSB), when $SU_L(3)\times U_N(1)$ is broken down to
$SU_L(2)\times U_Y(1)$, a pair of massive gauge bosons associated
with four broken generators of $SU_L(3)$ emerge in a doublet of
the electroweak group. Contrary to what happens in other theories,
the couplings between the new and the SM gauge bosons do not
involve any mixing angle, which means that they are expected to be
similar in magnitude to the ones existing between the SM gauge
bosons themselves.

Apart from the minimal 331 model, another version including
right-handed neutrinos has been considered in the literature more
recently \cite{Long,Long-1}. Its main feature is that it requires
a more economic Higgs sector to break the gauge symmetry and
generate the fermions masses. This model predicts the existence of
a singly-charged boson $Y^\pm$ along with a no self-conjugate
neutral boson $Y^{0*}$. Both of these new gauge bosons can be
classified as bileptons since they carry lepton number $L =\pm
\,2$, and thus are responsible for lepton-number violating
interactions \cite{Cuypers}. The neutral bilepton is a very
promising candidate in accelerator experiments since it may be the
source of neutrino oscillations \cite{Long-Inami}. The dynamical
behavior of the $Y^0$ boson is somewhat similar to that of the $W$
gauge boson, due to the nontrivial quantum number assignment. For
instance, the $Y^0 Y^+W^+$ coupling resembles those existing
between the electroweak gauge bosons. In the fermionic sector, the
$Y^0$ also couples to the quark pairs $(d,D_1)$, $(s,D_2)$, and
$(t,T)$, with $D_1$, $D_2$ and $T$ three new quarks predicted by
the model. These couplings induce nonzero static electromagnetic
properties for the neutral bilepton.

This presentation has been organized as follows. In Sec.
\ref{mod}, we present a brief review of the $331$ model with
right-handed neutrinos, with special emphasis on the current and
Yang-Mills sectors. Sec. \ref{cal} is devoted to the calculation
of the on-shell vertex $Y^0Y^{0*}\gamma$. In Sec. \ref{num} we
analyze the behavior of the $Y^0$ form factors, and  the
conclusions are presented in Sec. \ref{con}.

\section{The 331 model with right-handed neutrinos}
\label{mod}$331$ models are based on the $SU_C(3)\times
SU_L(3)\times U_N(1)$ gauge group. In the version with
right-handed neutrinos \cite{Long} the leptons are arranged as
\begin{equation}
f^i_L=\left( \begin{array}{ccc} \nu^i_L \\
e^i_L\\
(\nu^c_L)^i
\end{array}\right) \sim (1,3,-1/3), \ \  e^i_R \sim (1,1,-1),\ \
i=1,\,2,\,3,
\end{equation}
where $i$ stands for the family index. In the quark sector, a new
quark for each family is necessary. The first two quark families
transform as
\begin{equation}
{Q_a}_L=\left( \begin{array}{ccc} {d_a}_L \\
-{u_a}_L\\
{D_a}_L
\end{array}\right) \sim (3,\bar{3},0), \ \
{u_a}_R \sim (3,1,2/3),  \ \  {d_a}_R \sim (3,1,-1/3),\, \ {D_a}_R
\sim (3,1,-1/3),
\end{equation}
for $a=1,\,2$, whereas the third family transforms differently
\begin{equation}
{Q_3}_L=\left( \begin{array}{ccc} {u_3}_L \\
{d_3}_L\\
T_L
\end{array}\right) \sim (3,3,1/3),\ \
{u_3}_R\sim (3,1,2/3),\ \ {d_3}_R\sim (3,1,-1/3),\ \ T_R\sim
(3,1,2/3).
\end{equation}

As far as the scalar sector is concerned, only three triplets of
$SU_L(3)$ are required to achieve  the SSB mechanism:
\begin{eqnarray}
\chi=\left( \begin{array}{ccc} \chi^0 \\
\chi^-\\
\chi^{'0}
\end{array}\right)\sim (1,3,-1/3),\ \
\rho=\left( \begin{array}{ccc} \rho^+ \\
\rho^0\\
\rho^{'+}
\end{array}\right)\sim (1,3,2/3),\ \
\eta=\left( \begin{array}{ccc} \eta^0 \\
\eta^-\\
\eta^{'0}
\end{array}\right)\sim (1,3,-1/3).
\end{eqnarray}
In contrast, the minimal version requires three triplets and one
sextet. The vacuum expectation values $<\chi>^T=(0,0,w/\sqrt{2})$,
$<\rho>^T=(0,u/\sqrt{2},0)$, and $<\eta>^T=(v/\sqrt{2},0,0)$ yield
the following SSB pattern
\begin{eqnarray}
SU_C(3)\times SU_L(3)\times U_N(1)\ \
\begin{array}{c}
w\\ \to
\end{array}\ \
SU_C(3)\times SU_L(2)\times U_Y(1)\ \
\begin{array}{c}
u,\,v \\ \to
\end{array}\ \
 SU_C(3)\times U_e(1).
\end{eqnarray}
Notice that in order to break $SU_C(3)\times SU_L(3)\times U_N(1)$
into $SU_C(3)\times SU_L(2)\times U_Y(1)$, only the scalar triplet
$\chi$ is required. The covariant derivative in the triplet
representation is given by
\begin{equation}
{\cal
D}_\mu=\partial_\mu-ig\frac{\lambda^a}{2}A^a_\mu-ig_NN\frac{\lambda^9}{2}N_\mu,
\end{equation}
where $\lambda^9=2\,{\rm diag}\{1,1,1\}/3$, and $\lambda^a$
($a=1\cdots 8$) are the Gell-Mann matrices. The generators are
normalized as ${\rm Tr}(\lambda^a \lambda^b)=2\delta^{ab}$ and
${\rm Tr}(\lambda^9 \lambda^9)=2$. In the first stage of SSB, five
generators of $SU_L(3)$ along with the one associated with
$U_N(1)$ are broken, {\it i.e.} $\lambda^a <\chi>_0\neq 0$, for
$a=4, \ldots, 9$. The linear combination
$Y=(3\sqrt{2}N\lambda^9-\lambda^8)/\sqrt{3}$ annihilates the
vacuum and can be identified with the hypercharge operator. In
this stage the three exotic quarks and the gauge bosons associated
with the broken generators of the $331$ group $Y^0$, $Y^\pm$, and
$Z^\prime$ acquire mass. The exotic quarks have the same electric
charge as the SM quarks, namely, $Q_{D_{1,\,2}}=-1/3$ and
$Q_T=2/3$. As for the massive gauge bosons, both $Y^0$ and $Y^\pm$
are complex, whereas $Z^\prime$ is a real field with no quantum
numbers from the electroweak group.

At the Fermi scale, when $SU_C(3)\times SU_L(2)\times U_Y(1)$ is
broken down to $SU_C(3)\times U_e(1)$, the masses of the heavy
particles receive new contributions. The diagonalization of the
complete Higgs kinetic-energy sector leads to the following
mass-eigenstate fields:
\begin{eqnarray}
Y^{0(*)}_\mu &=&\frac{1}{\sqrt{2}}\left(A^4_\mu \mp iA^5_\mu\right), \\
Y^\mp_\mu&=&\frac{1}{\sqrt{2}}\left(A^6_\mu \mp iA^7_\mu\right),\\
W^\pm_\mu&=&\frac{1}{\sqrt{2}}\left(A^1_\mu \mp iA^2_\mu\right),
\end{eqnarray}
with $m^2_{Y^0}=g^2(w^2+u^2)/4$, $m^2_{Y^\pm}=g^2(w^2+v^2)/4$, and
$m^2_W=g^2(u^2+v^2)/4$. The symmetry-breaking hierarchy yields a
splitting between the bilepton masses:

\begin{equation}
\label{splitting} |m^2_{Y^0}-m^2_{Y^\pm}|\leq m^2_{W}.
\end{equation}

It is straightforward to obtain the explicit Lagrangian for the
current sector. We will concentrate only on those terms involving
the complex field $Y^0$, which in the lepton sector only couples
to neutrinos, whereas in the quark sector it couples to both SM
and exotic quarks as follows:
\begin{equation}
 \label{l1} {\cal
L}^{NC}_{Y^0}=\frac{g}{\sqrt{2}}\left(-\sum_{i=1,\,2}\bar{d}_{iL}\gamma^\mu
D_{iL}+\bar{u}_{3L}\gamma^\mu T_L\right)Y^0_\mu +{\rm H.c.}
\end{equation}
This is the only term of the fermion sector that contributes to
the one-loop induced $Y^0Y^{0*}\gamma$ vertex, whereas in the
bosonic sector there are contributions from both gauge and charged
scalar fields. In this work we will not consider those
contributions arising from the latter and concentrate only on the
Yang-Mills sector.

\subsection{ The Yang-Mills sector of 331 models}
In order to calculate the gauge-sector contributions to the
$Y^0Y^{0*}\gamma$ vertex, it is necessary to introduce the
gauge-fixing term. We found it convenient to use the unitary gauge
for our calculation. Since the Yang-Mills sector was discussed to
a certain extent in the case of the minimal version of the model
\cite{Tavares-331}, we refrain from presenting a more detailed
discussion and focus on those points relevant for the present
discussion. The Yang-Mills sector associated with the group
$SU_L(3)\times U_N(1)$ is given by
\begin{equation}
{\cal L}_{YM}=-\frac{1}{4}F^a_{\mu \nu}F^{\mu
\nu}_a-\frac{1}{2}N_{\mu \nu}N^{\mu \nu},
\end{equation}
where $F^a_{\mu \nu}=\partial_\mu A^a_\nu-\partial_\nu
A^a_\mu+f^{abc}A^b_\mu A^c_\nu$ and $N_{\mu \nu}=\partial_\mu
N_\nu-\partial_\nu N_\mu$, being $f^{abc}$ the structure constants
of the group $SU_L(3)$. We can write this Lagrangian as
\begin{equation}
\label{L_YM} {\cal L}_{YM}={\cal L}^{SM}_{YM}+{\cal
L}^{SM-NP}_{YM}+{\cal L}^{NP}_{YM},
\end{equation}
where the first term represents the Yang-Mills sector associated
with the electroweak group:
\begin{equation}
{\cal L}^{SM}_{YM}=-\frac{1}{4}F^i_{\mu \nu}F^{\mu
\nu}_i-\frac{1}{4}B_{\mu \nu}B^{\mu \nu},  \ \ \ \ \ i=1,\,2,\,3.
\end{equation}
The term ${\cal L}^{SM-NP}_{YM}$ represents the interactions
between the SM gauge fields and the heavy ones. It can be written
in the following $SU_L(2)\times U_Y(1)$-invariant form
\begin{eqnarray}
\label{15} {\cal L}^{SM-NP}_{YM}&=&-\frac{1}{2}\left(D_\mu
Y_\nu-D_\nu Y_\mu\right)^\dag \left(D^\mu Y^\nu-D^\nu
Y^\mu\right)-iY^\dag_\mu\left( g{\bf F}^{\mu
\nu} +g'{\bf B}^{\mu \nu}\right)Y_\nu \nonumber \\
&&-\frac{ig}{2}\frac{\sqrt{3-4s^2_W}}{c_W}Z^\prime_{\mu}\left(Y^\dag_\nu
\left(D^\mu Y^\nu-D^\nu Y^\mu\right)-\left(D^\mu Y^\nu-D^\nu
Y^\mu\right)^\dag Y_\nu\right),
\end{eqnarray}
where $Y^\dag_\mu=(Y^{0*}_\mu, Y^+_\mu)$ is a doublet of the
electroweak group with hypercharge $-1$ and
$D_\mu=\partial_\mu-ig{\bf A}_\mu +ig'{\bf B}_\mu$ is the
covariant derivative associated with this group. In addition, we
have introduced the definitions ${\bf F}_{\mu
\nu}=\sigma^iF^i_{\mu \nu}/2$, ${\bf A}_\mu=\sigma^iA^i_\mu/2$,
and ${\bf B}_\mu=YB_\mu/2$, with $\sigma^i$ the Pauli matrices.
Finally, the last term in Eq. (\ref{L_YM}) is also invariant under
the electroweak group and comprises the interactions between the
heavy gauge fields:
\begin{eqnarray}
{\cal L}^{NP}_{YM}&=&-\frac{1}{4}Z^\prime_{\mu \nu}{Z^\prime}^{\mu
\nu}+\frac{g^2}{4}\left(Y^\dag_\mu \frac{\sigma^i}{2}Y_\nu
-Y^\dag_\nu \frac{\sigma^i}{2}Y_\mu\right)\left(Y^{\dag \mu}
\frac{\sigma^i}{2}Y^\nu
-Y^{\dag \nu} \frac{\sigma^i}{2}Y^\mu\right)\nonumber \\
&&+\frac{3g^2}{16}\left(Y^\dag_\mu Y_\nu-Y^\dag_\nu
Y_\mu\right)\left(Y^{\dag \mu} Y^\nu-Y^{\dag \nu} Y^\mu
\right)-\frac{3g^2}{4}Z^\prime_{\mu}Y^\dag_\nu
\left(Z^\mu_2 Y^\nu-Z^\nu_2 Y^\mu \right)\nonumber \\
&&-\frac{ig}{2}\frac{\sqrt{3-4s^2_W}}{c_W}\,Y^\dag_\mu
Y_\nu\,{Z^\prime}^{\mu \nu}.
\end{eqnarray}

From these Lagrangians we have derived the Feynman rules shown in
Table \ref{tab1}, which are necessary for the calculation of the
gauge boson contribution to the $Y^0Y^{0*}\gamma$ vertex. These
results are in agreement with Ref. \cite{Long-1}

\begin{table}[!hbt]
\begin{tabular}{rr}
\hline Vertex&Feynman rule\\
\hline\hline $Y^0_\alpha
(p)W^-_\lambda (k_1)Y^+_\rho (k_2)$& $ ig\left((p-k_2)_\lambda
g_{\rho \alpha}+(k_2-k_1)_\alpha g_{\lambda \rho}+(k_1-p)_\rho
g_{\alpha
\lambda}\right)/\sqrt{2}$\\
$A_\mu (q)V^+_\lambda(k_1) V^-_\rho (k_2)$& $
-ie\left((k_2-k_1)_\mu g_{\lambda \rho}+(q-k_2)_\lambda g_{\mu
\rho} +(k_1-q)_\rho g_{\mu \lambda}\right)$
\\
$Y^0_\alpha Y^{0*}_\beta Y^+_\lambda Y^-_\rho$& $
ig^2\left(2g_{\alpha \rho}g_{\beta \lambda}-g_{\alpha
\lambda}g_{\beta \rho}-g_{\alpha \beta}g_{\lambda \rho}\right)/2$
\\
$Y^0_\alpha Y^{0*}_\beta W^+_\lambda W^-_\rho$& $
ig^2\left(2g_{\alpha \lambda}g_{\beta \rho}-g_{\alpha
\beta}g_{\lambda \rho}-g_{\alpha \rho}g_{\beta \lambda}\right)/2$
\\
$A_\mu Y^0_\alpha Y^+_\lambda W^-_\rho$& $ ige\left(g_{\alpha
\lambda}g_{\rho \mu}-2g_{\alpha \mu}g_{\lambda \rho}+g_{\alpha
\rho}g_{\lambda \mu}\right)/\sqrt{2}$
\\ \hline\hline
\end{tabular}
\caption{\label{tab1}Feynman rules necessary for the calculation
of the gauge boson contribution to the $Y^0Y^{0*}\gamma$ vertex.
$V^\pm$ stands for $Y^\pm$ or $W^\pm$. All the 4-momenta are
directed inward.}
\end{table}

\section{The static electromagnetic properties of the $Y^0$ boson}
\label{cal} We turn now to the calculation of the static
electromagnetic properties of the no self-conjugate neutral boson
$Y^0$.  In the usual notation, the most general on-shell
$Y^0_\alpha Y^0_\beta A_\mu$ vertex can be written as
\cite{Hagiwara,nieves}
\begin{equation}
\label{elecprop} \Gamma_{\alpha \beta \mu}=i\,e\left(2\Delta
\kappa (q_\beta g_{\alpha \mu}-q_\alpha g_{\beta
\mu})+\frac{4\,\Delta Q}{m^2_{Y^0}}\,p_\mu q_\alpha
q_\beta+2\Delta \widetilde{\kappa}\,\epsilon_{\alpha \beta \mu
\lambda}q^\lambda+\frac{4\,\Delta
\widetilde{Q}}{m^2_{Y^0}}\,q_\beta \epsilon_{\alpha \mu \lambda
\rho}p^\lambda q^\rho\right).
\end{equation}
Note that the $p_\mu g_{\alpha \beta}$ term, which is present for
a charged particle, is absent as it would violate gauge
invariance. This term can only arise through the electromagnetic
covariant derivative. The magnetic (electric) dipole moment
$\mu_{Y^0}$ ($\widetilde {\mu}_{Y^0}$) and the electric (magnetic)
quadrupole moment $Q_{Y^0}$ ($\widetilde{Q}_{Y^0}$) are given in
terms of the electromagnetic form factors as follows
\begin{eqnarray}
\label{mu}
\mu_{Y^0}&=&\frac{e}{2\,m_{Y^0}}(2+\Delta \kappa), \\
\label{Q}
Q_{Y^0}&=&-\frac{e}{m^2_{Y^0}}(1+\Delta \kappa+\Delta Q),\\
\label{mutilde}
\widetilde{\mu}_{Y^0}&=&\frac{e}{2\,m_{Y^0}}\Delta \widetilde{\kappa},\\
\label{Qtilde} \widetilde{Q}_{Y^0}&=&-\frac{e}{m^2_{Y^0}}(\Delta
\widetilde{\kappa}+\Delta \widetilde{Q}).
\end{eqnarray}

The CP-violating form factors $\Delta \widetilde{\kappa}$ and
$\Delta \widetilde{Q}$ are not induced in the $331$ model with
right-handed neutrinos. In the fermionic sector, $\Delta
\widetilde \kappa$ can be induced at the one-loop level, but it
requires that the neutral boson couples to both left- and
right-handed fermions simultaneously \cite{Burgess,TTU}.

In order to compute the contributions to the on-shell
$Y^0Y^{0*}\gamma$ vertex, we used the method described in Refs.
\cite{Tavares-331,Stuart}, which is a generalization of the
Passarino-Veltman reduction scheme \cite{Passarino}. Since the
gauge invariant form (\ref{elecprop}) is obtained once all the
contributions are summed over, the absence of the $p_\mu g_{\alpha
\beta}$ term and the cancellation of ultraviolet divergences will
serve as a test to check the correctness of our results. Below we
will present separately the fermionic and gauge boson
contributions to the $\Delta Q$ and $\Delta \kappa$ form factors.

\subsection{Fermion contribution}
\begin{figure}
\centering
\includegraphics[width=3.5in]{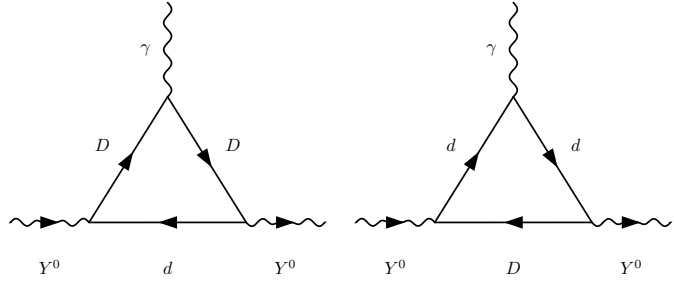}
\caption{\label{Fermion-diag} Feynman diagrams for the fermion
contributions to the static quantities of the $Y^0$ boson.}
\end{figure}

The contribution of this sector comes from the Feynman diagrams
shown in Fig. \ref{Fermion-diag}. There are two triangle diagrams
for each quark pair $(d$, $D_1)$, $(s$, $D_2)$, and $(t$, $T)$. We
will denote by $q$ the SM quark and by $q^\prime$ the exotic one.
Once the reduction scheme described above is applied to solve the
loop amplitudes, the contribution from the ($q$, $q^\prime$) quark
pair can be written as

\begin{eqnarray}
\label{DQ_fer} \Delta Q^{\rm Ferm.}&=& 6\,a\,Q\Bigg\{
\frac{2}{\Delta_{qq^\prime}}\,\left( x_{q^\prime} - x_q \right)
\,\left( 1
-3\,\left(x_{q^\prime}+x_q\right)+2\,\left(x_{q^\prime}-x_q\right)^2\right)
\,{\rm arccosh}\left(\frac{x_{q^\prime}  + x_q
-1}{2\,{\sqrt{x_{q^\prime} \,x_q
}}}\right)\nonumber\\&+&4\left(x_{q^\prime} -x_q\right)  + \left(
x_{q^\prime}+x_q-2\,\left(x_{q^\prime}-x_q\right)^2  \right)
\,\log \left(\frac{x_{q^\prime} }{x_q }\right) \Bigg\},
\end{eqnarray}
\begin{eqnarray}
\label{Dk_fer} \Delta \kappa^{\rm Ferm.}&=&9\,a\,Q\,\left(
x_{q^\prime}  - x_q \right) \, \Bigg\{
\frac{2}{\Delta_{q{q^\prime}} }\,\left(x_{q^\prime} + x_q
-\left(x_{q^\prime}-x_q\right)^2 \right) \, {\rm
arccosh}\left(\frac{x_{q^\prime} + x_q-1}{2\,{\sqrt{x_{q^\prime}
\,x_q }}}\right)\nonumber\\ &-& 2 + \left( x_{q^\prime} - x_q
\right) \, \log \left(\frac{x_{q^\prime} }{x_q }\right) \Bigg\},
\end{eqnarray}

\noindent with $a=g^2/(96\,\pi^2)$, $x_i=m_i^2/m_{Y^0}^2$ and
$\Delta_{ij}^2=(x_i+x_j-1)^2-4\,x_j\,x_j$. A factor of 3 has been
included to account for the quark color number, and $Q$ stands for
the quark charge in units of that of the positron. Eqs.
(\ref{DQ_fer}) and (\ref{Dk_fer}) are to be summed over the $(d$,
$D_1)$, $(s$, $D_2)$, and $(t$, $T)$ quark pairs.

Both $\Delta Q^{\rm Ferm.}$ and $\Delta \kappa^{\rm Ferm.}$ are
antisymmetric under the interchange of $x_q$ and $x_{q'}$, which
means that they vanish when the $q$ and $q'$ quarks are
degenerate. Since it is expected that the exotic quarks are
heavier than the SM ones ($x_{q'}\gg x_{q}$), it would be
interesting to have analytical expressions for the scenario in
which $x_q\sim 0$ and $x_{q^\prime}$ is arbitrary. After some
algebra, Eqs. (\ref{DQ_fer}) and (\ref{Dk_fer}) yield

\begin{eqnarray}
\Delta Q^{\rm Ferm.}&=&12\,a\,Q\,x_{q^\prime}
\left(2+\left(2\,x_{q^\prime}-1\right)
\log\left(\frac{|x_{q^\prime}-1|}{x_{q^\prime}}\right)\right),
\\
\Delta \kappa^{\rm
Ferm.}&=&18\,a\,Q\,x_{q^\prime}\left(1+x_{q^\prime}
\log\left(\frac{|x_{q^\prime}-1|}{x_{q^\prime}}\right)\right).
\end{eqnarray}

In the heavy-mass limit, $\Delta \kappa^{\rm Ferm.}\to -9\,a\,Q$
and $\Delta Q^{\rm Ferm.}\to 0$. Of course when $x_{q^\prime}\to
0$, the degenerate fermion case is recovered and both form factors
vanish.

\subsection{Gauge boson contribution}

\begin{figure}
\centering
\includegraphics[width=3.5in]{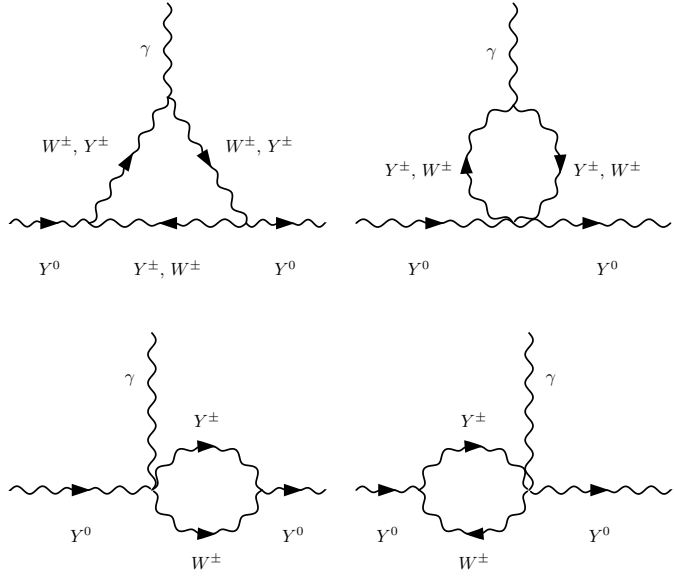}
\caption{\label{Boson-diag}Feynman diagrams for the gauge boson
contributions to the static quantities of the $Y^0$ boson.}
\end{figure}

We found it convenient to make the calculation for this
contribution in the unitary gauge. Although the triangle diagrams
give rise to fourth-order tensor integrals due to the longitudinal
part of the gauge boson propagators, our calculation scheme is
suited to work out this class of terms straightforwardly. The
static electromagnetic properties of the $Y^0$ boson arise from
the six Feynman diagrams shown in Fig. \ref{Boson-diag}, whose
amplitudes can be constructed out of the Feynman rules presented
in Table \ref{tab1}. After solving the loop integrals, the full
amplitude can be cast in the form of Eq. (\ref{elecprop}), which
leads to

\begin{eqnarray}
\label{DQ_bos}\Delta Q^{\rm Bos.}&=& \frac{\,a}{2\,x_Y \,x_W
}\,\left(\Delta_{YW}^2+ 12\,x_X\,x_W\right)
\,\Bigg\{4\,\left(x_Y-x_W\right)+4\,
\left(\left(x_Y+x_W\right)-2\,\left(x_Y -x_W\right)^2\right)
\,\log \left(\frac{x_Y }{x_W }\right) \nonumber\\
&+& \frac{2}{\Delta_{YW}}\,\left( x_Y  - x_W \right) \, \left( 1 -
3\left(\,x_W+x_Y\right) +2\,\left(x_Y-x_W\right)^2 \right) \, {\rm
arccosh}\left(\frac{ x_Y + x_W -1}{2\,{\sqrt{x_Y \,x_W
}}}\right)\Bigg\},
\end{eqnarray}
and
\begin{eqnarray}
\label{Dk_bos} \Delta \kappa^{\rm Bos.}&=& \frac{3\,a}{2\,x_Y
\,x_W }\,\Bigg\{\left( x_Y - x_W \right)
\,\left(1+\left(x_Y-x_W\right)^2 - 2\,\left(x_Y +x_W-6\,x_Y\,x_W
\right)  \right) \nonumber\\&-&
\Big(x_Y\,\left(1-x_Y\right)^2\,\left(3+x_Y\right)+
x_Y\,x_W\,\left(x_Y\,\left(8\,x_Y-9\,x_W-13\right)+9\right)\nonumber\\&+&
x_W\,\left(1-x_W\right)^2\,\left(3+x_W\right)+x_Y\,x_W\,\left(x_W\,\left(8\,x_W-9\,x_Y-13\right)+9\right)
\Big) \, \log\left(\frac{x_Y }{x_W }\right) \nonumber\\ &-&
2\,\left( x_Y - x_W \right) \,{\Delta_{YW}} \left( 3
-\left(x_Y-x_W\right)^2 + 2\,\left(x_Y +x_W+6\,x_Y\,x_W \right)
\right) \,{\rm arccosh}\left(\frac{x_Y  + x_W-1 }{2\,{\sqrt{x_Y
\,x_W }}}\right) \Bigg\},
\end{eqnarray}
with $x_W=m_W/m_{Y^0}$ and $x_{Y}=m_{Y^\pm}/m_{Y^0}$. Due to the
mass splitting (\ref{splitting}), the bileptons would be nearly
degenerate if $m_{Y^\pm}\ge m_W$. Therefore it is worth obtaining
analytical expressions for the form factors in this scenario. Eqs.
(\ref{DQ_bos}) and (\ref{Dk_bos}) yield the following results for
$x_Y=1$:

\begin{eqnarray}
\label{DQ_bos-1}\Delta Q^{\rm Bos.}&=& \frac{a}{2} \,\left( 8 +
x_W\right)\,\Bigg\{ \, 4\,\left(1-x_W\right)+\left( 1 + x_W
\,\left(2\,x_W-5  \right)  \right)\,\log(x_W)\nonumber\\
&+&\frac{2}{{\sqrt{\left(x_W-4  \right) \,x_W }}}\,\left( 1 - x_W
\right) \,x_W \, \left(2\,x_W -7 \right) {\rm
arccosh}\left(\frac{\sqrt{x_W}}{2}\right) \Bigg\},
\end{eqnarray}
and
\begin{eqnarray}
\label{Dk_bos-1} \Delta \kappa^{\rm Bos.}&=&
\frac{3\,a}{4}\Bigg\{2\,\left(1 - x_W  \right) \,\left( 8 + x_W
\right) +\left(16 + \left( x_W-3  \right) \,x_W \,\left( 12 + x_W
\right)\right)\,\log(x_W)\nonumber\\ &-&2\,\left(x_W -1 \right)
\,{\sqrt{\left(x_W-4  \right) \,x_W }}\, \left( 12 + x_W
\right){\rm arccosh}\left(\frac{\sqrt{x_W}}{2}\right) \Bigg\}.
\end{eqnarray}

From the previous results, it is easy to see that the
contributions to the $Y^0$ form factors are antisymmetric under
the interchange of the masses of the particles circulating in the
loop, which means that they vanish when these particles are
degenerate, i.e. $m_q=m_{q'}$ and $m_W=m_Y$.

\section{Numerical evaluation}
\label{num}

We turn now to the numerical analysis of the $Y^0$ form factors.
We would like to emphasize that our main aim is to estimate the
size and behavior of the form factors in some illustrative
scenarios rather than making a careful study of the allowed
parameter space of the model, which is beyond the present work.

In addition to the mass of the $Y^0$ boson, there are four other
unknown parameters which enter into the $Y^0$ form factors. These
are the masses of the three exotic quarks $m_{D_1}$, $m_{D_2}$,
and $m_T$, together with the charged bilepton mass $m_{Y^\pm}$.
Since the splitting between the bilepton masses is bounded, i.e.
$|m^2_{Y^0}-m^2_{Y^\pm}|\leq m^2_{W^+}$, $m_{Y^\pm}$ is bounded
once $m_{Y^0}$ is fixed. Although in the minimal 331 model the
bilepton masses are bounded from above at 1 TeV as a result of
matching the gauge couplings constants at the Fermi scale, which
leads to $\sin \theta_W \le 1/4$ \cite{Ng}, in the version with
right-handed neutrinos the same condition leads to $\sin \theta_W
\le 3/4$, which yields less stringent constraints on the bilepton
masses. The most recent bounds indicate that $m_{Y^0}$ is greater
than 100 GeV \cite{Long,Bounds}. We will thus analyze the form
factors in the range 100 GeV $\le m_{Y^0}\le$ 500 GeV.

As for the exotic quarks, although there are bounds on the masses
of the exotic quarks predicted in other SM extensions, to our
knowledge there are no such bounds in the specific case of the 331
model with right-handed neutrinos. However, it is reasonable to
assume that the exotic quarks are heavier than the top quark.
Therefore, for the corresponding masses we will consider values
ranging from 200 to 800 GeV. Furthermore, as will be shown below,
the maximal value of the fermionic contribution to the static
quantities of the $Y^0$ boson is reached in this mass range. Below
we will evaluate separately the fermion and boson contribution to
the $Y^0$ form factors.

\subsection{Fermion contribution}

\begin{figure}
\centering
\includegraphics[width=2.5in]{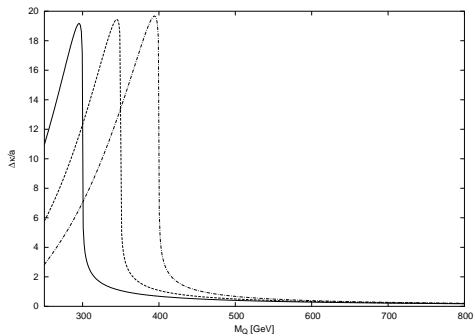}
\caption{\label{DK-fer-fig1}The fermion contribution to the
$\Delta \kappa$ form factor as a function of the mass of the
exotic quarks, which are assumed to be degenerate, for different
values of the neutral bilepton mass: 300 (continuous line), 350
(dashes), and 400 GeV (dashes and points).}
\end{figure}

\begin{figure}
\centering
\includegraphics[width=2.5in]{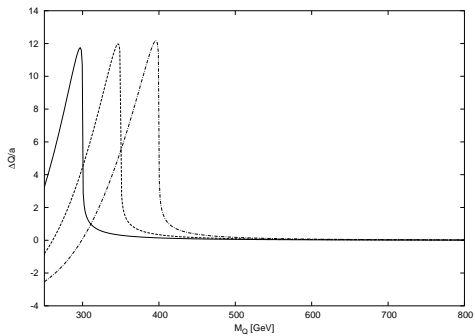}
\caption{\label{DQ-fer-fig1}The same as in Fig. \ref{DK-fer-fig1}
for the $\Delta Q$ form factor.}
\end{figure}

The general behavior of the fermion contribution to the static
quantities of the charged $W$ boson has been discussed to a large
extent in the literature \cite{Burgess,TTU,WWg-SM,WWg-NP}. The
main peculiarity of the CP-even electromagnetic form factors of a
neutral particle is that the contribution arising from a
degenerate fermion pair vanishes since the amplitude is
antisymmetrical under the interchange $m_q \to m_{q^\prime}$.
Although the latter is also true for an arbitrarily charged gauge
boson, their CP-even static quantities do not vanish for
degenerate fermions since $Q_q \ne Q_{q^\prime}$. In the following
analysis we will consider the scenario in which the exotic quarks
are degenerate, with a mass $m_Q$. As already explained, we will
consider the range 200 GeV$\leq m_Q \leq $ 800 GeV. In Figs.
\ref{DK-fer-fig1} and \ref{DQ-fer-fig1} we show the $\Delta
\kappa$ and $\Delta Q$ form factors as a function of $m_Q$ for
some illustrative values of the neutral bilepton mass $m_{Y^0}$,
namely, 300, 350 and 400 GeV. We note that the curves displayed in
Figs. \ref{DK-fer-fig1} and \ref{DQ-fer-fig1} are the full
contribution from the three quark families. In the range under
consideration for $m_Q$, the form factors are considerable smaller
for $m_{Y^0} \le $ 200 GeV. We can clearly observe that there is a
dramatic enhancement in the $m_{Y^0}$ threshold $m_{Y^0}=m_q+m_Q$,
which stems from the fact that the respective quark pair ($q,\,Q$)
can be directly produced from the bilepton provided that
$m_{Y^0}\ge m_q+m_Q$. Above the threshold and in the heavy mass
limit, both form factors decrease rapidly and vanish when $m_Q$ is
much larger than the mass of the SM quarks. It is interesting to
point out that  the individual contributions to $\Delta \kappa$
from each fermion pair tend to the constant value $-9\,a$ in the
heavy fermion limit, whereas $\Delta Q$ vanishes. This is in
accordance with the decoupling theorem \cite{Appelquist}: since
$\Delta \kappa$ is associated with a term that arises from
dimension-four operators, it is expected to be sensitive to
nondecoupling effects of heavy physics, whereas $\Delta Q$ cannot
be sensitive to this class of effects as it is associated with a
term generated by a nonrenormalizable dimension-six operator
\cite{Inami}. In spite of the nondecoupling nature of the
contributions from each quark family, the full $\Delta \kappa$
vanishes in the heavy fermion limit. It turns out that the partial
contributions, which are proportional to the quark charge, become
constant and their sum vanishes since it is proportional to
$Q_{D_1}+Q_{D_2}+Q_T=0$. This is to be contrasted with the
behavior of the fermion contribution to the $\Delta \kappa$ form
factor of the $W$ boson in the heavy fermion limit. In this case
the contribution of each quark family is proportional to
$Q_u-Q_d=1$, thus the sum over the three quark families does not
vanish.

From Figs. \ref{DK-fer-fig1} and \ref{DQ-fer-fig1} we can conclude
that $\Delta \kappa $ can be of the order of $10\,a$, whereas
$\Delta Q$ is about one order of magnitude below. This behavior is
similar to that observed for the size of the fermion contribution
to the electromagnetic form factors of the $W$ boson in the SM
\cite{WWg-SM} and some of its extensions \cite{WWg-NP}. Although
the maximal value of the form factors is reached around the
threshold $m_{Y^0}= m_q+m_Q$, there is no reason to expect that
such a scenario is realized in nature. The scenarios shown through
Figs. \ref{DK-fer-fig1} and \ref{DQ-fer-fig1} are very
illustrative of the behavior of the quark contribution to the
static quantities of the $Y^0$ boson and so we refrain from
presenting the most general case in which the exotic quark are
nondegenerate.

\subsection{Gauge boson contribution}

\begin{figure}
\centering
\includegraphics[width=2.5in]{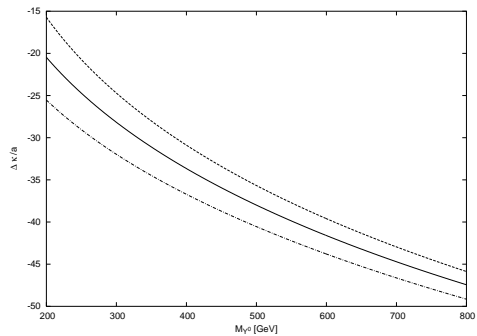}
\caption{\label{DK-bos-fig1}The gauge boson contribution to the
$\Delta \kappa$ form factor of the $Y^0$ boson as a function of
its mass when $m_{Y^\pm}=m_{Y^0}$ (continuous line),
$m_{Y^\pm}^2=m_{Y^0}^2-m_W^2$ (dashes) and
$m_{Y^\pm}^2=m_{Y^0}^2+m_W^2$ (dashes and points). The last two
curves correspond to the case when the $m_{Y^\pm}$ reaches its
maximal and minimal allowed values. $\Delta \kappa$ is restricted
to lie in the strip bounded by the extremal lines.}
\end{figure}

\begin{figure}
\centering
\includegraphics[width=2.5in]{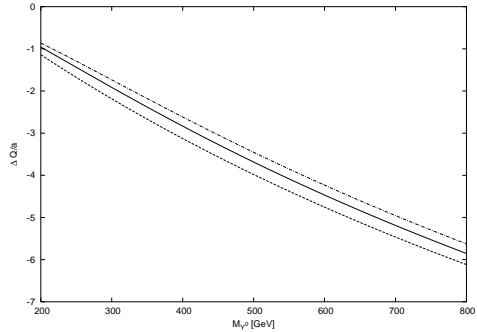}
\caption{\label{DQ-bos-fig1}The same as in Fig. \ref{DK-bos-fig1}
for the $\Delta Q$ form factor.}
\end{figure}

In Figs. \ref{DK-bos-fig1} and \ref{DQ-bos-fig1} we show the
contributions from the gauge bosons to the electromagnetic form
factors of the $Y^0$ boson as a function of $m_{Y^0}$ when the
bileptons are degenerate and also when $m_{Y^\pm}$ reaches its
minimal and maximal allowed values: $m_{Y^\pm}^2=m_{Y^0}^2-m_W^2$
and $m_{Y^\pm}^2=m_{Y^0}^2+m_W^2$. The form factors are restricted
to lie in the strip bounded by the extremal lines. Although the
form factors seem to increase indefinitely as $m_{Y^0}$ increases,
they tend to a constant value for very large $m_{Y^0}$. There is
no contradiction with the decoupling limit as one cannot make
large the internal mass $m_{Y^\pm}$ while keeping fixed the
external mass $m_{Y^0}$ due to the bound (\ref{splitting}).
Furthermore, the quantities which have physical meaning are the
magnetic dipole and electric quadrupole moments [See Eqs.
(\ref{mu}) and (\ref{Q})], which do vanish for very large
$m_{Y^0}$. From Figs. \ref{DK-bos-fig1} and \ref{DQ-bos-fig1}, it
is evident that $\Delta \kappa$ is one order of magnitude larger
than $\Delta Q$ for each value of $m_{Y^0}$. The fact that the
size of  $\Delta \kappa$ is larger than that of $\Delta Q$ has
been also observed for the case of the electromagnetic form
factors of the charged $W$ boson form within all of the theories
studied up to now.

To obtain the total contribution to the $Y^0$ form factors, it is
necessary to sum over the fermion and gauge boson contributions,
along with the one arising from the scalar sector of the theory.
Apart from the specific details of the model, we do not expect
that the size of the scalar contribution is different to that
observed in the case of the $W$ form factors. In that case, the
scalar sector yields a marginal correction. In fact a very large
number of Higgs bosons would be required to yield a large
correction.

\section{Summary}
\label{con}

A neutral vector boson can have static electromagnetic properties
provided that the associated field is no self-conjugate. We have
presented the calculation of the static electromagnetic properties
of the neutral no self-conjugate boson $Y^0$ which arises in the
$SU(3)_c\times SU(3)_L \times U(1)_N$ model with right-handed
neutrinos. This model is interesting since it requires that the
fermion families be a multiple of the quark color number in order
to cancel anomalies, thereby suggesting a solution to the family
problem. It has been pointed out that the $Y^0$ boson is a good
candidate in high energy experiments since it may be the source of
neutrino oscillations as it is responsible of lepton-number
violating interactions. The calculation was done in the unitary
gauge and the fermion and gauge boson contributions were obtained
by a modified version of the Passarino-Veltman reduction scheme.
As a crosscheck, the form factors were obtained independently by
the Feynman parameter technique and the results, expressed in
terms of parametric integrals, were numerically evaluated and
compared with the results obtained via the Passarino-Veltman
method. A perfect agreement was observed. In this model the $Y^0$
boson only couples to left-handed fermions and so only the CP-even
form factors are induced at the one-loop level. The behavior of
both contributions was analyzed. In the fermion sector there is
the contribution of the three quark pairs $(D_1,\, d)$, $(D_2,\,
s)$, and $(T,\,t)$, with $D_1$, $D_2$, and $T$ three exotic quarks
whose charge is identical to that of the respective SM quark. As
for the gauge boson contribution, there is the contribution of a
singly charged bilepton $Y^\pm$. The symmetry breaking hierarchy
yields an upper bound on the splitting between the bilepton masses
such that $|m_{Y^0}^2-m_{Y^\pm}^2|\le m_W^2$, which means that the
bileptons are nearly degenerate provided that their mass is
heavier than $m_W$. From the numerical analysis we can conclude
that the size of the $Y^0$ form factors is somewhat similar to
that observed for the $W$ boson form factors in the SM and some of
its extensions.

\acknowledgments{Support from CONACYT and SNI  is acknowledged.
G.T.V. also acknowledges partial support from SEP-PROMEP.}

\end{document}